\documentclass[journal]{IEEEtran}
\usepackage{cite}
\usepackage{amsfonts}
\usepackage{textcomp}
\usepackage{epstopdf}
\usepackage{tabu}
\usepackage{textcomp}
\usepackage{mathtools}
\usepackage{amssymb}
\usepackage{nccmath}
\usepackage{lipsum}
\usepackage{cuted, nccmath}
\usepackage{multicol}
\usepackage{ragged2e}
\usepackage{amsmath}
\usepackage{algorithmic}
\usepackage{array}
\usepackage{stfloats}
\usepackage{url}
\usepackage{epstopdf}
\usepackage{tabu}
\usepackage{graphicx}
\usepackage{caption}
\usepackage{flushend}

\newcommand{\overbar}[1]{\mkern 1.5mu\overline{\mkern-1.5mu#1\mkern-1.5mu}\mkern 1.5mu}
\makeatletter
\newcommand{\mathleft}{\@fleqntrue\@mathmargin0pt}
\newcommand{\mathcenter}{\@fleqnfalse}
\makeatother

\usepackage[printwatermark]{xwatermark}
\usepackage{xcolor}
\ifCLASSINFOpdf
\else
\fi
\usepackage{textcomp}
\usepackage{epstopdf}
\usepackage{tabu}
\usepackage{textcomp}
\usepackage{mathtools}
\usepackage{amssymb}
\usepackage{nccmath}
\usepackage{lipsum}
\usepackage{cuted, nccmath}
\usepackage{multicol}
\usepackage{ragged2e}
\usepackage{amsmath}

\usepackage{array}

\usepackage{stfloats}
\usepackage{url}

\begin{document}
%
\title{Performance of Two-Way Relaying over $\alpha$-$\mu$ Fading Channels in Hybrid RF/FSO Wireless Networks}
%
%
%

\author{\normalsize Mohammed~A. Amer,
        and Suhail~Al-Dharrab~\\ 
        \small King Fahd University of Petroleum and Minerals,\\
        \small  Department of Electrical Engineering, Saudi Arabia\\
        \small E-mail: \{g201605360, suhaild\}@kfupm.edu.sa}
\maketitle

\begin{abstract}
In this paper, the performance of two-way relaying in mixed RF/FSO communication system with a backup RF link is investigated. Uplink RF channels are used to send data of $K$ users to a two-way relay, $R$, whereas FSO link is mainly used to exchange data between a base station $S$ and $R$. We propose to have a backup RF link between the relay $R$ and the node $S$ to improve reliability under certain conditions. All uplink RF channels follow Rayleigh fading model while $\alpha$-$\mu$ is adopted to model both backup RF and FSO links. We approximate the widely used Gamma-Gamma fading model using the $\alpha$-$\mu$ distribution based on moments-based estimator technique assuming perfect alignment between transmitter and receiver antennas. This approximation shows good agreement under certain atmospheric turbulence conditions. Then, we derive exact closed-form expressions for the outage and average symbol error probabilities and derive approximations at high signal-to-noise ratio (SNR). We corroborate our analytical expressions with extensive Monte-Carlo simulations and demonstrate exact match. Furthermore, we analyze the effect of number of nodes,  opportunistic scheduling among $K$ nodes, and $\alpha$-$\mu$ parameters on the overall performance of mixed RF/FSO and backup RF systems. Our numerical results illustrate an achievable coding gain when increasing $K$; however, performance degradation occurs as the relay applies selection that favors the domination of specific links in the system.
\end{abstract}


%
\IEEEpeerreviewmaketitle

\section{Introduction}
%
%
%
%
\IEEEPARstart{T}{here} has been an increasing demand on higher data rates over the last years with our current congested spectrum. New technologies have emerged, which require efficient utilization of power and bandwidth as possible in designing the future wireless communication systems. For instance, cooperative relaying have been introduced lately to fulfill the quality of service (QoS) requirements for upcoming wireless networks. Specifically, in wireless dense networks, relays can improve coverage area, reliability, and enhance the spectral efficiency when deployed in wireless networks. Therefore, it could be a key feature in the next generation (5G) deployment of wireless cellular networks~\cite{0}.\par

Due to the inflation in number of wireless devices operating in the Radio Frequency (RF) bands, and expected increase of their applications in the near future, the RF spectrum is congested and the cost of the licensed band in RF spectrum is relatively high. Therefore, there has been proposals for alternative technologies to complement the current RF spectrum such as free-space optical (FSO) communication systems. Wireless FSO systems have gained high attention among researchers for their features such as higher capacity and higher achievable data rates compared to the RF based systems. While many RF links can be licensed, the FSO links are free-band  which result to have less cost at the moment~\cite{0}. Recently, mixed RF/FSO in wireless communication systems has numerous advantages specially that it can be deployed as a backhaul structure/link to increase the reliability and speed of the backbone network. This will meet the requirements of higher data rates and QoS.\par

The performance analysis of dual-hop one-way mixed RF/FSO links was analyzed in~\cite{2}. Lee \textit{et al.} have investigated the system over Gamma-Gamma fading model and under the assumption of perfect pointing error. They derived the end-to-end outage probability. In \cite{3}, Ansari \textit{et al.} studied the impact of pointing error on the performance over the same system model. The case of two-way, both half-duplex and full duplex, relaying scheme was studied in~\cite{4,5}. Authors in \cite{4,5} have derived the end-to-end outage probability and average symbol error probability.\par

There have been studies on the case of multiple wireless nodes in the RF link part of mixed RF/FSO system. In~\cite{6,7}, Miridakis \textit{et al.} and Salhab \textit{et al.} investigated one-way relaying multi-node networks with Decode-and-Forward (DF) and fixed gain Amplify-and-Forward (AF) relaying in mixed RF/FSO. They assumed Gamma-Gamma fading model in their derivations of end-to-end outage and average symbol error probabilities. Al-Eryani \textit{et al.} in~\cite{8} proposed an order selection for the best SNR wireless node with bi-directional half-duplex communications in mixed RF/FSO assuming Gamma-Gamma fading model. Different fading channel models have been adopted for the FSO link such as Gamma-Gamma, M\'{a}laga, inverse-$K$, and log-normal Rician fading distributions~\cite{10,5}.\par
 Due to the unexpected behavior in atmospheric turbulence and possible pointing alignment error, the hybrid RF/FSO system can encounter significant performance degradation; which is not counted in~\cite{8}. Therefore, a backup RF link is essential at the two-way relay to prevent the communication loss in the event of outage in the FSO link, or performing below the required signal-to-noise ratio (SNR).\par

To the best of our knowledge, no research results on the two-way relaying in dual-hop system over mixed RF/FSO with an independent backup RF link considering $\alpha$-$\mu$ fading model. We also propose a generalized opportunistic scheduling/selection scheme, and derive the outage and average symbol error probabilities for the aforementioned system. The main contributions of this work can be summarized as follows:\par
\begin{itemize}
\item Propose a backup RF link in the multi-node dual-hop system over hybrid RF/FSO based on generic channel fading model using a generalized opportunistic scheduling selection scheme.
\item  Provide an approximation to the widely used Gamma-Gamma fading model for atmospheric turbulence using the proposed $\alpha$-$\mu$ distribution.
\item Derive accurate closed-form analytical expressions for both outage and average symbol error probabilities, and asymptotic approximations for high SNR.
\item Analyze the effect of number of nodes, opportunistic scheduling, and $\alpha$-$\mu$ parameters on the overall performance of mixed RF/FSO and backup RF/RF systems.
\end{itemize}\par
The rest of the paper is organized as follows. In Section II, we describe the system model and its associated channel models as well as the cumulative distribution function (CDF) derivation of each link. The transmission protocol, and adopted opportunistic scheduling in RF and mixed RF/FSO are in Section III. The exact analysis of outage probability of the system is derived in Section IV. Section V covers the average symbol error probability derivation. In Section VI, we obtain the asymptotic expressions of the outage probability for high SNR. In Section VII, we present and discuss extensive numerical simulation results. Finally, we conclude the paper in Section VIII.


\section{System and Channel Model}
We consider a system of a base station node $S$, a single two-way DF relay $R$, and $K$ nodes/sensors $N_1,\dots,N_K$. Base station $S$ supports both FSO and RF communications with the two-way relay $R$, which has both optical and RF transceivers. In FSO link, the transmitter uses subcarrier intensity modulation technique~\cite{10}, and the node $R$ along with the nodes $N_1,\dots,N_K$ are connected by RF links as shown in Fig. 1.

 First, we will investigate RF links between the $k^{th}$ node and two-way relay, i.e. $N_k \rightarrow R$ and $R \rightarrow N_k$, $k = 1,\dots,K$. The received signal at the relay node $R$ from the $k^{th}$ node $N_k$ is given by
\begin{equation}
\begin{aligned}
r_{R}^{RF}=\sqrt{P_{N_{k}}}g_{N_{k},R}x_{N_{k}}+n_{R},
\end{aligned}
\end{equation}
while the received signal at the node $N_k$ is
\begin{equation}
\begin{aligned}
r_{N_{k}}^{RF}=\sqrt{P_{R}}g_{R,N_{k}}x_{R}+n_{N_{k}}, \quad  k = 1,\dots,K,
\end{aligned}
\end{equation}
where $P_{N_{k}}$ and $P_{R}$ denote the average electrical signal power of $N_k$ and $R$ nodes, respectively. $g_{N_{k},R}$ and $g_{R,N_{k}}$ are the small-scale fading coefficients over $N_k \rightarrow R$ and $R \rightarrow N_k$ links. $x_{N_{k}}$ and $x_{R}$ denote the $N_k$ node and relay transmitted symbols with $\mathbb{E}\big[|x_{i}|^{2}\big]=1, i\in \{N_{k},R\}$, respectively, and $\mathbb{E}[.]$ stands for the expectation. $n_{R}$ and $n_{N_{k}}$ represent the zero-mean additive white Gaussian noise at the relay $R$ and the node $N_k$ with power spectral density (PSD) of ${N_{0,R}^{RF}}$ and ${N_{0,k}^{RF}}$. The instantaneous SNRs at the input of the relay $R$ and the $k^{th}$ node $N_k$ are respectively given by 
\begin{equation}
\gamma_{N_{k},R}=\frac{P_{N_{k}}}{N_{0,R}^{RF}} |g_{N_{k},R}|^2,
\end{equation}
\begin{equation}
\gamma_{R,N_k}=\frac{P_{R}}{{N_{0,k}^{RF}}} |g_{R,N_{k}}|^2.
\end{equation}
On the other hand, the received optical signal at the relay $R$ and base station $S$ are given by~\cite{2}
\begin{equation}
r_{R}^{OP}=h_{S,R}\bigg\{\sqrt{P_{S}^{OP}}(1+M x_{S})\bigg\}+n_{R}^{OP},
\end{equation}
\begin{equation}
r_{S}^{OP}=h_{R,S}\bigg\{\sqrt{P_R^{OP}}(1+M \hat{x}_{R})\bigg\}+n_{S}^{OP},
\end{equation}
where $h_{R,S}$ and $h_{S,R}$ represent the channel fading coefficients of the $R$ $\rightarrow$ $S$ and $S$ $\rightarrow$ $R$ wireless FSO links, respectively. $P_{S}^{OP}$ and $P_{R}^{OP}$ stand for the average optical power of the transmitted symbols of base station $S$ and relay $R$, respectively. $M$ represents the modulation index,  $x_{S}$ is the transmitted symbol by $S$, and $\hat{x}_{R}$ is the decoded symbol at the relay $R$ where $\mathbb{E}\big[|x_{S}|^{2}\big]=\mathbb{E}\big[|\hat{x}_{R}|^2\big]=1$. The AWGN terms at the input of the relay $R$ and base station $S$ are denoted by $n_{R}^{OP}$ and $n_{S}^{OP}$ with zero mean and PSDs of $N_{0,R}^{OP}$ and $N_{0,S}^{OP}$, respectively. The relation between optical and RF electrical power is given by $P_{R}^{OP}=\xi_{1}^{R} P_{R}$ and $P_{S}^{OP}=\xi_{1}^{S} P_{S}$ where $\xi_{1}^{R}$ and $\xi_{1}^{S}$ stand for the electrical-to-optical conversion ratios~\cite{2}. However, the optical-to-electrical conversion ratios at the relay $R$ and base station $S$ are given by $\xi_{2}^{R}$ and $\xi_{2}^{S}$, respectively.
The instantaneous SNR at input of the relay $R$ and $S$ are
\begin{equation}
\begin{aligned}
\gamma_{S,R}=\frac{\xi_{1}^{S}\xi_{2}^{R}P_{S}}{N_{0,R}^{OP}}|h_{S,R}|^2,
\end{aligned}
\end{equation}
\begin{equation}
\begin{aligned}
\gamma_{R,S}=\frac{\xi_{1}^{R}\xi_{2}^{S}P_{R}}{N_{0,S}^{OP}}|h_{R,S}|^2.
\end{aligned}
\end{equation}

\begin{figure}[!t]
\centering
\includegraphics[scale=0.35]{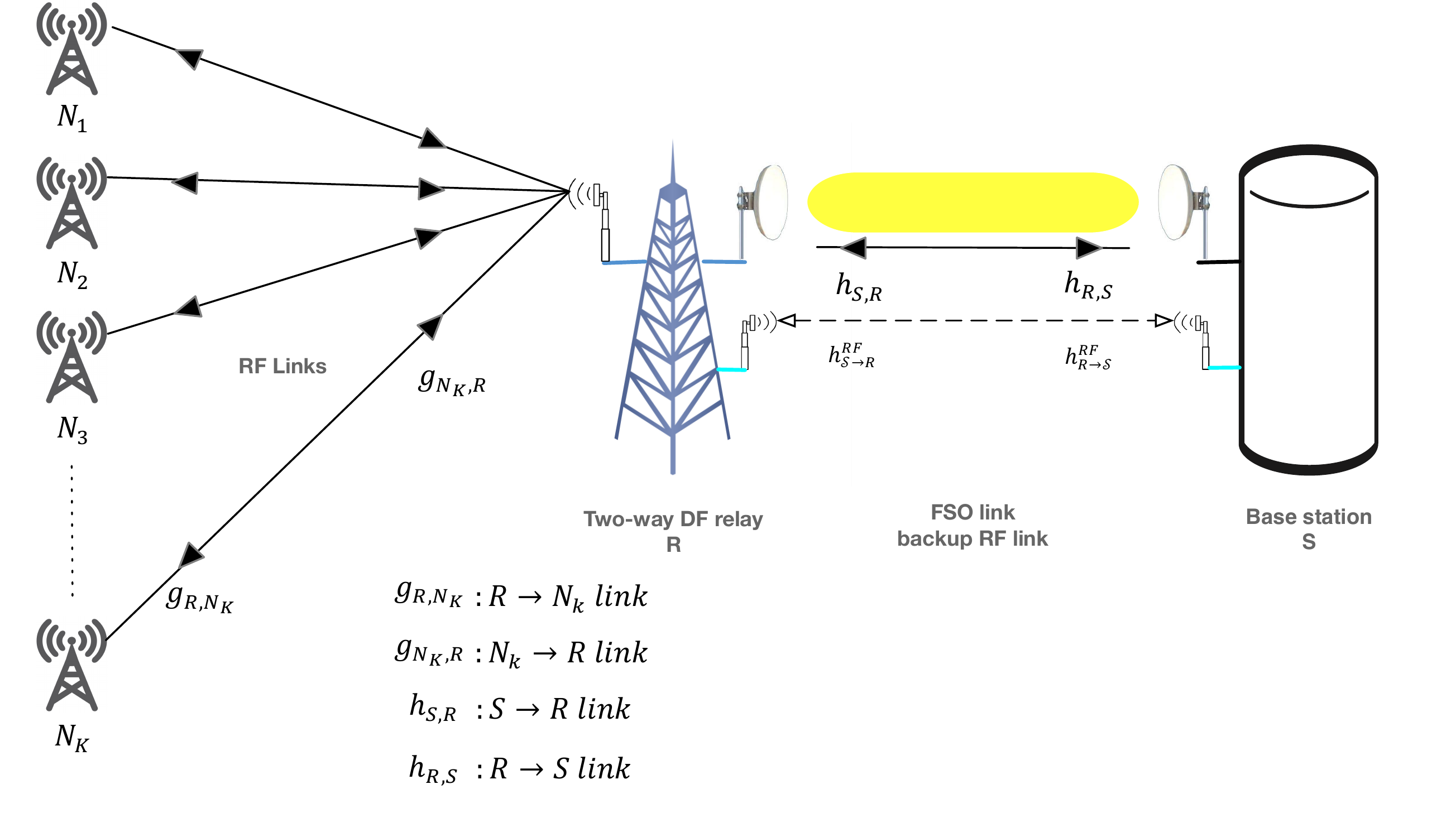}
\caption{Wireless network with two-way relaying over hybrid FSO/RF links.}
\label{Fig.1}
\end{figure}\par

\subsection{RF Channel Model}\par

We assume channel coefficients $g_{R,N_{k}}$ and $g_{N_{k},R}$, for all $ k = 1,\dots,K$ follow the Rayleigh fading model in RF links. This implies that $|g_{R,N_{k}}|^{2}$ and $|g_{N_{k},R}|^{2}$ are exponentially distributed, i.e. chi-square distributed with two degrees of freedom, which is generally given by~\cite{11}
\begin{equation}
\begin{aligned}
f_{\gamma_{X,Y}}(\gamma_{x,y})=\frac{1}{\overbar{\gamma}_{X,Y}}e^{-{\frac{\gamma_{x,y}}{\overbar{\gamma}_{X,Y}}}}, \quad \gamma_{x,y}>0
\end{aligned}
\end{equation}
where $\gamma_{X,Y}$ represents the SNR for $N_{k}\to R$ (or $R\to N_{k}$) links, and $\overbar{\gamma}_{X,Y}=\mathbb{E}\big[\gamma_{X,Y}\big]$ is the average received SNR. The CDF of the SNR in the RF link is $ F_{\gamma_{X,Y}}{(\gamma_{x,y})}=1-e^{-{\frac{\gamma_{x,y}}{\overbar{\gamma}_{X,Y}}}} ,\gamma_{x,y}>0$.

\subsection{Hybrid FSO and RF Channels}\par

We consider a primary FSO transmission between the base station $S$ and the relay $R$ with an independent secondary/backup RF link. As weather conditions can affect severely the FSO link reliability, we propose to have a hybrid communication system that replaces FSO with RF in these scenarios. In this paper, we assume channel coefficients $h_{S,R}$ and $h_{R,S}$ are modeled by the generalized $\alpha$-$\mu$ small-scale fading model contrary to many previous works which assume Gamma-Gamma distribution for the FSO channel~\cite{8}. The probability density function (PDF) of the $\alpha$-$\mu$ generalized fading model for $X\rightarrow Y$ link is given by~\cite{12}
\begin{equation}
\begin{aligned}
\small f_{h_{X,Y}}{(h_{x,y})}=\frac{{\alpha}{\mu^{\mu}}{(h_{x,y})^{\alpha\mu-1}}}{\Gamma{(\mu)}{\Omega_{\alpha}^{\alpha\mu}}}e^{-\mu {(\frac{h_{x,y}}{\Omega^{\alpha}}})^{\alpha}} ,{\mu\geq0},{\alpha\geq0},\\ {h_{x,y}\geq0},
\end{aligned}
\end{equation}
where $\alpha$ and $\mu$ are the power and fading parameters, respectively. $\Gamma(.)$ is the complete Gamma function, and the parameter $\Omega_{\alpha}=\sqrt[\alpha]{\mathbb{E}\big[|h_{R,S}|^{\alpha}\big]}$. The distribution of instantaneous SNR for FSO link considering $\alpha$-$\mu$ fading channel is~\cite{13}
\begin{equation}
\begin{aligned}
f_{\gamma_{X,Y}}(\gamma_{x,y})=\frac{\alpha}{2\Gamma{(\mu)}}({\frac{\mu}{(\overbar{\gamma}_{x,y})^{\frac{\alpha}{2}}}})^{\mu}({\gamma_{x,y}})^{\frac{\alpha\mu}{2}-1}e^{-\mu({\frac{\gamma_{x,y}}{\overbar{\gamma}_{x,y}}})^{\frac{\alpha}{2}}},
\\{\mu\geq0}, {\alpha\geq0},  {\gamma_{x,y}\geq0},
\end{aligned}
\end{equation}
where $\gamma_{X,Y}$ represents the SNR for $S\to R$ (or $R\to S$) links, and $\overbar{\gamma}_{x,y}$ is the average received SNR. The CDF of the $\alpha$-$\mu$ fading channel is given by
\begin{equation}
F_{\gamma_{X,Y}}(\gamma_{x,y})=\frac{\gamma_{inc} ({\mu,\mu({\frac{\gamma_{x,y}}{\overbar{\gamma}_{x,y}}})^{\frac{\alpha}{2}})}}{\Gamma({\mu})},
\end{equation}
where $\gamma_{inc} {(.,.)}$ is the lower incomplete Gamma function~\cite{14}. In general, the $\alpha$-$\mu$ fading distribution is used to model the non-linear propagation in addition to the multipath propagation through arbitrary medium based on the physical parameters $\alpha$ and $\mu$. As a special case, Nakagami-$m$, Rayleigh, One-sided Gaussian, Exponential, and Weibull distributions can be derived from the $\alpha$-$\mu$ fading distribution using $\alpha$ and $\mu$ as in Table I~\cite{12}.
\begin{table}[!t]
\small\addtolength{\tabcolsep}{19pt}
\caption{Derived PDFs from the $\alpha$-$\mu$ Fading Distribution.}
\label{Table.1}
\centering
\begin{tabular}{ l c c } 
\hline
$f_{X}{(x)}$ & $\alpha$ & $\mu$\\ [0.5ex]
\hline
One-sided Gaussian & 2 & 0.5\\
Rayleigh & 2 & 1\\
Weibull  & 1.75 & 1\\
Nakagami-$m$ & 2 & 2\\
Exponential & 1 & 1\\
\hline
\end{tabular}
\end{table}

\subsection{Approximation of Gamma-Gamma Distribution}\par

In this section, we approximate independent and identically distributed (i.i.d.) Gamma-Gamma fading model using the generic $\alpha$-$\mu$ distribution by moment-based estimators. The $\alpha$-$\mu$ distribution has less complexity compared to Gamma-Gamma, which simplifies the analysis of FSO systems and preserves accurate results. The $n^{th}$ moment of Gamma-Gamma random variable $X$ is given by~\cite{18}
\begin{equation}
\begin{aligned}
\mathbb{E}[X^{n}]=(\eta\beta)^{-n}\frac{\Gamma{(\eta+n)}\Gamma{(\beta+n)}}{\Gamma{(\eta)}\Gamma{(\beta)}},
\end{aligned}
\end{equation}
where $\eta$ and $\beta$ are the Gamma-Gamma atmospheric turbulence parameters, and the $n^{th}$ moment of the $\alpha$-$\mu$ fading distribution is given by~\cite{12}
\begin{equation}
\begin{aligned}
\mathbb{E}[X^{n}]=\overbar{\varrho}^{-n}\frac{\Gamma{(\mu+\frac{n}{\alpha})}}{\mu^{\frac{n}{\alpha}}\Gamma{(\mu)}},
\end{aligned}
\end{equation}
where $\overbar{\varrho}$ stands for the $\alpha^{th}$ root mean value of envelope. By equating the first, second, and third moments of both fading channels and numerically solving for unknown variables, we obtain
\begin{equation}
\begin{aligned}
&\mathbb{E}[X]=\overbar{\varrho}^{-1}\frac{\Gamma{(\mu+\frac{1}{\alpha})}}{\mu^{\frac{1}{\alpha}}\Gamma{(\mu)}}=(\eta\beta)^{-1}\frac{\Gamma{(\eta+1)}\Gamma{(\beta+1)}}{\Gamma{(\eta)}\Gamma{(\beta)}}\\
&\mathbb{E}[X^{2}]=\overbar{\varrho}^{-2}\frac{\Gamma{(\mu+\frac{2}{\alpha})}}{\mu^{\frac{2}{\alpha}}\Gamma{(\mu)}}=(\eta\beta)^{-2}\frac{\Gamma{(\eta+2)}\Gamma{(\beta+2)}}{\Gamma{(\eta)}\Gamma{(\beta)}}\\
&\mathbb{E}[X^{3}]=\overbar{\varrho}^{-3}\frac{\Gamma{(\mu+\frac{3}{\alpha})}}{\mu^{\frac{3}{\alpha}}\Gamma{(\mu)}}=(\eta\beta)^{-3}\frac{\Gamma{(\eta+3)}\Gamma{(\beta+3)}}{\Gamma{(\eta)}\Gamma{(\beta)}},
\end{aligned}
\end{equation}
where values of $\eta$ and $\beta$ for weak and strong atmospheric turbulence conditions and their corresponding $\alpha$ and $\mu$ are given in Table II. Fig. 2 depicts the PDF comparison of the Gamma-Gamma distribution with the approximated $\alpha$-$\mu$ for specific atmospheric turbulence conditions. We notice a good match between the two PDFs for very weak ($\eta=21.5,\beta=19.8$) and weak (a) ($\eta=9.70,\beta=8.20$) atmospheric turbulence conditions. However, for severe atmospheric turbulence conditions ($\eta=4 \,\, \text{or}\,\, 4.34,\beta=1.84 \,\,\text{or}\,\, 1.30$), the approximation mismatch the Gamma-Gamma PDF, and higher order moments are needed to improve the approximation.

\begin{table}[!t]
\caption{Atmospheric Turbulence Conditions, and $\alpha$-$\mu$ Parameters.}
\label{Table.2}
\centering
\small\addtolength{\tabcolsep}{2pt}
\begin{tabular}{ c c c c c c} 
\hline
Atmospheric Turbulence & $\eta$ & $\beta$ & $\alpha$ & $\mu$ \\ [0.5ex]
\hline
Very Weak & 21.5 & 19.8 & 2.34 & 2.21 \\
Weak (a) & 9.70 & 8.2 & 1.68 & 1.85 \\
Weak (b) & 8.65 & 7.14 & 2 & 1.3695 \\
Severe (a) & 4 & 1.84 & 0.537 & 2.022\\
Severe (b) & 4.34 & 1.30 & 0.579 & 2.723\\
\hline
\end{tabular}
\end{table}

\begin{figure}[!t]
\centering
\includegraphics[scale=0.45]{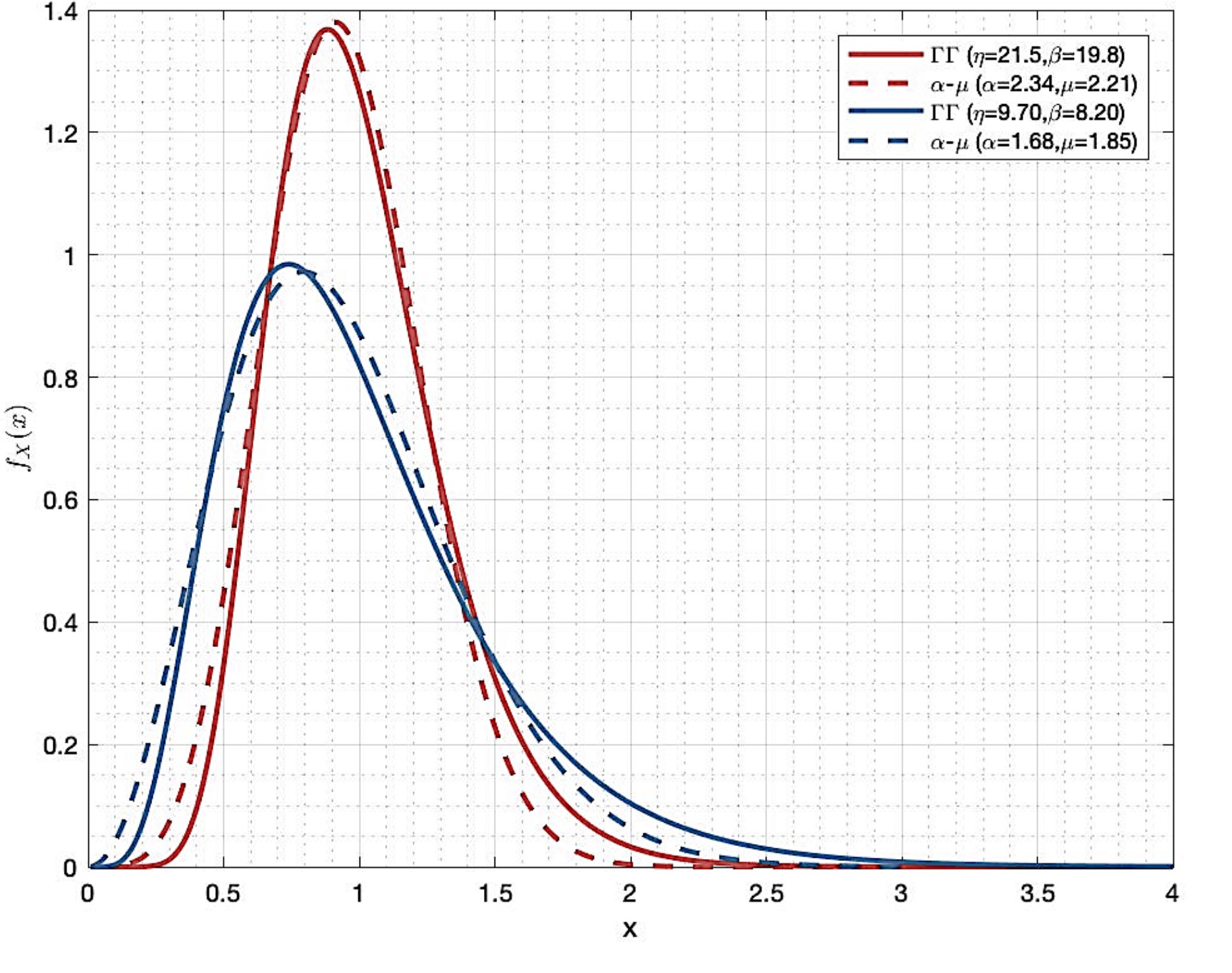}
\caption{The $\alpha$-$\mu$ approximation of Gamma-Gamma distribution for specific atmospheric turbulence conditions under perfect antennas alignment.}
\label{Fig.2}
\end{figure}

\section{Transmission Protocol}\par
Transmission between the base station $S$ and $k^{th}$ wireless node $N_k$ is achieved through the relay $R$ in two phases. In the first phase, denoted by $ST^{(1)}$, both base station $S$ and $k^{th}$ node $N_k$ transmit their optical and RF signals to relay $R$. While in the second phase, denoted by $ST^{(2)}$, the relay $R$ converts the RF received signal of $N_k$ to an optical one, and transmits it to the base station $S$. At the same time, it converts the optical signal received from base station $S$ to RF to be transmitted to the $k^{th}$ selected wireless node based on an opportunistic scheduling scheme that will be discussed in the next subsection.

\subsection{Opportunistic Scheduling in RF links}\par
In the RF links, we propose an opportunistic scheduling scheme among $K$ nodes, i.e. $N_1,\dots,N_K$, where the relay $R$ will select the best SNR among them. The CDF of the selected node, $M^* \in \{N_1,\dots,N_K\}$, by the relay $R$ is given by
\begin{equation}
\begin{aligned}[c]
F_{\gamma_{M^*,R}}(\gamma)&=Pr\{{\max({\gamma_{N_{1},R},\dots,\gamma_{N_{K},R})\leq{\gamma_{th}}}\}} \\
&=Pr\{{{\gamma_{N_{1},R}\leq{\gamma_{th}},\dots,{\gamma_{N_{K},R}\leq{\gamma_{th}}}}}\},
\end{aligned}
\end{equation}
where $\gamma_{M^*,R}=\max(\gamma_{N_{1},R},\dots,\gamma_{N_{K},R})$, and $Pr\{.\}$ denotes the probability operation. Consider all RF channels to be independent and identically distributed (i.i.d.), then the joint CDF of the overall SNR can be obtained as
\begin{equation}
\begin{aligned}
F_{\gamma_{M^{*},R}}(\gamma)={\prod_{k=1}^{K}F_{\gamma_{N_{k},R}}(\gamma)}.
\end{aligned}
\end{equation}
Upon substituting the CDF of $\gamma_{N_{k},R}$ in (17), the resulting CDF is given by
\begin{equation}
\begin{aligned}
F_{\gamma_{M^{*},R}}(\gamma)=\bigg({1-e^{-{\frac{\gamma}{\overbar{\gamma}_{n^{*},r}}}}}\bigg)^{K}, \gamma >0
\end{aligned}
\end{equation}
where $\overbar{\gamma}_{n^{*},r} = \mathbb{E}\big[\gamma_{N_k,R}\big], k=1,\dots,K$. Let $F_G(\gamma)={1-e^{-{\frac{\gamma}{\overbar{\gamma}_{n^{*},r}}}}}$, then we differentiate (18) in order to obtain the PDF of the SNR at selected node as
\begin{equation}
\begin{aligned}
f_{\gamma_{M^{*},R}}(\gamma)=K\bigg(F_G(\gamma)\bigg)^{K-1}\frac{dF_G(\gamma)}{d\gamma}, \gamma >0
\end{aligned}
\end{equation}
Using the general Binomial relation, the CDF of the best SNR is given by
\begin{equation}
\begin{aligned}
F_{\gamma_{M^{*},R}}(\gamma)=K \sum_{k=0}^{K-1} {K-1\choose k}{\frac{(-1)^{k}}{(k+1)}}{(1-e^{-{\frac{(k+1)\gamma}{\overbar{\gamma}_{n^{*},r}}}}}).
\end{aligned}
\end{equation}

In case the node with best SNR and/or others are unavailable due to scheduling, we consider an extension of (20) to a generalized order selection scheme. 

The SNRs among available $N < K$ nodes will form an ordered SNR set, and $N$ is the order of selected node. Relay selects the highest SNR among them, i.e. the $N^{th}$ best SNR denoted by $N^*$. The CDF of the selected node with SNR $\gamma_{N^*,R}$ is given by 
\begin{equation}
\begin{aligned}
F_{\gamma_{N^*,R}}(\gamma)=&K{K-1\choose N-1} \sum_{k=0}^{K-N} {K-N\choose k}{\frac{(-1)^{k}}{(k+N)}}\\
&\times{(1-e^{-{\frac{(k+N)}{\overbar{\gamma}_{n^{*},r}}}\gamma}}).
\end{aligned}
\end{equation}

On other hand, the CDF of the downlink ($R \rightarrow N^*$), $F_{\gamma_{R,N^*}}(\gamma)$, is given by 
\begin{equation}
\begin{aligned}
F_{\gamma_{R,N^*}}{(\gamma)}=1-e^{-{\frac{\gamma}{\overbar{\gamma}_{r,n^{*}}}}}.
\end{aligned}
\end{equation}
where $\overbar{\gamma}_{r,n^{*}} = \mathbb{E}\big[\gamma_{R,N_k}\big], \forall\,\, k\in \{1,\dots,K\}$.

\subsection{Opportunistic Scheduling in Mixed RF/FSO links}\par

\vspace*{10pt}
In the RF/FSO and backup RF/RF systems, the CDF of the bi-directional links $S\to R$ or $R \to S$ is written in terms of Meijer's G-function using~\cite{17}[Sec. 8.4.16/Eq.1] as 
\begin{equation}
\begin{aligned}
F_{\gamma_{X,Y}}^{FSO/RF}{(\gamma_{x,y})}=\frac{1}{\Gamma{(\mu)}}G_{1,2}^{1,1}\left[\mu({\frac{\gamma_{x,y}}{\overbar{\gamma}_{x,y}}})^{\frac{\alpha}{2}}\left|\begin{matrix}1\\
\mu,0\end{matrix}\right.\right]
\end{aligned},
\end{equation}
where $\gamma_{X,Y}$ represents the SNR for $S\to R$ (or $R\to S$) links, and $\large{G}_{c,d}^{a,b} \left[ \cdot  \left|
\begin{smallmatrix} .,. \\ .,. \end{smallmatrix} \right. \right]$ is the Meijer's G-function defined in~\cite{14}. The notation $FSO/RF$ refers to the CDF of $\alpha$-$\mu$ fading channel for both backup RF and FSO links.

\section{Exact Outage Probability Analysis}

\vspace*{10pt}
The outage probability analysis is essential to characterize the error performance and reliability. In two-way relaying system, stability of bi-directional transmission links is critical compared to the one-way case as nodes must be active during the two transmission phases. 

The outage occurs when one of the two phases experiences an outage event, which takes place when the SNR of any link, e.g. FSO or RF links, drops below a predetermined threshold value, $\gamma_{th}$, i.e. $P_{out}=Pr\{\gamma\leq\gamma_{th}\}$. Consequentially, total outage probability in two-way relaying is obtained by
\begin{equation}
\begin{aligned}
P_{out}^{Tot}\triangleq ST_{out}^{(1)}+ST_{out}^{(2)}-ST_{out}^{(1)}ST_{out}^{(2)},
\end{aligned}
\end{equation}
where $ST_{out}^{(i)}, i=1,2$ stands for the outage event in the $i^{th}$ transmission phase.  So, the outage probability for $ST_{out}^{(1)}$ is defined by~\cite{5}
\begin{equation}
\begin{aligned}
ST_{out}^{(1)}&\triangleq Pr{\{\min{(\gamma_{N^*,R},\gamma_{S,R})}\leq{\gamma_{th}}\}}\\
&=1-Pr{\{\gamma_{N^*,R}>\gamma_{th}, \gamma_{S,R}>\gamma_{th}\}}.
\end{aligned}
\end{equation}
Since the two links are i.i.d., we can rewrite (25) as
\begin{equation}
\begin{aligned}
ST_{out}^{(1)} =& F_{\gamma_{N^*,R}}^{}{(\gamma_{th})}+F_{\gamma_{S,R}}^{FSO/RF}{(\gamma_{th})}\\
&- F_{\gamma_{N^*,R}}^{}{(\gamma_{th})}F_{\gamma_{S,R}}^{FSO/RF}{(\gamma_{th})}.
\end{aligned}
\end{equation}

By substituting (21), (23) in (26) we get the expression in (27) (shown at the top of next page).
\newpage
\begin{strip}
\begin{equation}
\begin{aligned}
ST_{out}^{(1)}=& K {K-1\choose N-1} \sum_{k=0}^{K-N} {K-N\choose k}{\frac{(-1)^{k}{(1-e^{-{\frac{(k+N)\gamma_{th}}{\overbar{\gamma}_{n^{*},r}}}}})}{(k+N)}}\bigg\{1-\frac{1}{\Gamma{(\mu)}}G_{1,2}^{1,1}\left[\mu({\frac{\gamma_{th}}{\overbar{\gamma}_{s,r}}})^{\frac{\alpha}{2}}\left|\begin{matrix}1\\
\mu,0\end{matrix}\right.\right]\bigg\}\\
&+\frac{1}{\Gamma{(\mu)}}G_{1,2}^{1,1}\left[\mu({\frac{\gamma_{th}}{\overbar{\gamma}_{s,r}}})^{\frac{\alpha}{2}}\left|\begin{matrix}1\\
\mu,0\end{matrix}\right.\right].
\end{aligned}
\end{equation}
\noindent\makebox[\linewidth]{\rule{18cm}{0.4pt}}
\end{strip}
Similarly, the definition of the outage probability for $ST_{out}^{(2)}$, where relay $R$ transmits the data to both base station $S$ and the selected node $N^*$, is given by
\begin{equation}
\begin{aligned}
ST_{out}^{(2)}&\triangleq Pr{\{\min{(\gamma_{R,N^*},\gamma_{R,S})}\leq{\gamma_{th}}\}}\\
&= 1-Pr{\{\gamma_{R,N^*}>\gamma_{th}, \gamma_{R,S}>\gamma_{th}\}}.
\end{aligned}
\end{equation}
As the two links are assumed i.i.d., we can rewrite (28) as
\begin{equation}
\begin{aligned}
ST_{out}^{(2)} =& F_{\gamma_{R,N^*}}^{}{(\gamma_{th})}+F_{\gamma_{R,S}}^{FSO/RF}{(\gamma_{th})}\\&- F_{\gamma_{R,N^*}}^{}{(\gamma_{th})}F_{\gamma_{R,S}}^{FSO/RF}{(\gamma_{th})}.
\end{aligned}
\end{equation}
Upon substituting (22), (23) in (29) we get
\begin{equation}
\begin{aligned}
ST_{out}^{(2)}=&~1-e^{-{\frac{1}{\overbar{\gamma}_{r,n^{*}}}}\gamma_{th}}\Bigg\{1-\frac{1}{\Gamma{(\mu)}}
G_{1,2}^{1,1}\left[\mu({\frac{\gamma_{th}}{\overbar{\gamma}_{r,s}}})^{\frac{\alpha}{2}}\left|\begin{matrix}1\\
\mu,0\end{matrix}\right.\right] \Bigg\}.
\end{aligned}
\end{equation}
Finally, by substituting (27) and (30) into (24) we obtain the outage probability for both hybrid RF/FSO and backup RF/RF systems in closed-from expression in (31).\begin{strip}
\noindent\makebox[\linewidth]{\rule{18cm}{0.4pt}}
\begin{equation}
\begin{aligned}
P_{out}^{Tot}=& K {K-1\choose N-1} \sum_{k=0}^{K-N} {K-N\choose k}{\frac{(-1)^{k}{(1-e^{-{\frac{(k+N)\gamma_{th}}{\overbar{\gamma}_{n^{*},r}}}}})}{(k+N)}}
\bigg\{1-\frac{1}{\Gamma{(\mu)}}G_{1,2}^{1,1}\left[\mu({\frac{\gamma_{th}}{\overbar{\gamma}_{s,r}}})^{\frac{\alpha}{2}}\left|\begin{matrix}1\\
\mu,0\end{matrix}\right.\right]\bigg\}\\
&+\frac{1}{\Gamma{(\mu)}}
G_{1,2}^{1,1}\left[\mu({\frac{\gamma_{th}}{\overbar{\gamma}_{s,r}}})^{\frac{\alpha}{2}}\left|\begin{matrix}1\\
\mu,0\end{matrix}\right.\right] \ + (1-e^{-{\frac{\gamma_{th}}{\overbar{\gamma}_{r,n^{*}}}}})\bigg\{1-\frac{1}{\Gamma{(\mu)}}
G_{1,2}^{1,1}\left[\mu({\frac{\gamma_{th}}{\overbar{\gamma}_{r,s}}})^{\frac{\alpha}{2}}\left|\begin{matrix}1\\
\mu,0\end{matrix}\right.\right] \bigg\}\\
&\times \bigg\{1-K {K-1\choose N-1} \sum_{k=0}^{K-N} {K-N\choose k}{\frac{(-1)^{k}{(1-e^{-{\frac{(k+N)\gamma_{th}}{\overbar{\gamma}_{n^{*},r}}}}})}{(k+N)}}\bigg\{1-\frac{1}{\Gamma{(\mu)}}G_{1,2}^{1,1}\left[\mu({\frac{\gamma_{th}}{\overbar{\gamma}_{s,r}}})^{\frac{\alpha}{2}}\left|\begin{matrix}1\\
\mu,0\end{matrix}\right.\right]\bigg\}\bigg\}.
\end{aligned}
\end{equation}
\end{strip}

\vspace*{-18pt}
\section{Average Symbol Error Probability Analysis}

We investigate and derive the average symbol error probability (ASEP) that reflects the reliability of a communication system under various environment conditions. In order to analyze the ASEP, CDF-based approach is adopted in our analysis after replacing $\gamma_{th}$ with $\gamma$. The CDF-based approach is given by~\cite{19}\\
\begin{equation}
\begin{aligned}
ASEP_{tot}= \frac{a\sqrt{b}}{2\sqrt{\pi}}\int_{0}^{\infty}\frac{e^{-b\gamma}}{\sqrt{\gamma}}F_{\gamma}^{tot}{(\gamma)}d\gamma,
\end{aligned}
\end{equation}
where parameters $a$ and $b$ are related to the modulation scheme in use and $a,b >0$. 

Sub-carrier intensity modulation (SIM) scheme is adopted in the system model and hence binary phase shift keying (BPSK) modulation can be used in both RF and FSO/RF links. Upon using (31), (7.813.1) and (3.381.4) in~\cite{14} and \cite{20} with straightforward mathematical manipulations, we obtain the $ASEP_{tot}$ for both systems in closed-form expression in (34) (given at the top of next page) where $G_{.,.:.,.:.,.}^{.,.:.,.:.,.}\left[\begin{matrix}.\cr .\end{matrix}\vert\begin{matrix}\ .,.\cr .,.\end{matrix}\vert\begin{matrix}\ .,.\cr .,.\end{matrix}\vert \psi,\chi \right]$ is the Extended Generalized  Bivariate Meijer's G-function (EGBMGF)~\cite{21}, $\Lambda_{1}=(b+\frac{1}{\overbar{\gamma}_{r,n_{k}}})$, and $\Lambda_{2}=\frac{(k+N+b\overbar{\gamma}_{n_{k},r}){\overbar{\gamma}_{r,n_{k}}+\overbar{\gamma}_{n_{k},r}}}{{\overbar{\gamma}_{r,n_{k}}\overbar{\gamma}_{n_{k},r}}}$. 

\vspace*{20pt}
\section{Asymptotic Analysis of Outage Probability}

The derived outage probability expression in (31) is complicated; hence, a simpler expression is needed to have insights on coding gain and diversity order of the system. At high SNR, the outage probability expression, in general, is approximated $P_{out}^{Tot}\simeq G_{c}({SNR})^{-G_{d}}$ where $G_{c}$ and $G_{d}$ stand for the coding gain and diversity order, respectively~\cite{11}. Assuming all channels are i.i.d. such that $\overbar{\gamma}_{n^*,r}=\overbar{\gamma}_{r,n^*}=\overbar{\gamma}_{s,r}=\overbar{\gamma}_{r,s}=\overbar{\gamma}_{\psi}$, we can further simplify (24) due to the fact that the product of multiple CDFs is insignificant compared other terms. The CDF of the end-to-end system can be rewritten by summing dominant CDFs in all three terms, which is given by 
\begin{equation}
\begin{aligned}
P_{out}^{\infty}\approx\lim_{\gamma\to\infty}\bigg\{&F_{\gamma_{N^{*},R}}^{}{(\gamma)}+F_{\gamma_{S,R}}^{FSO/RF}{(\gamma)}\\&+F_{\gamma_{R,N^{*}}}^{}{(\gamma)}+F_{\gamma_{R,S}}^{FSO/RF}{(\gamma)}\bigg\}.
\end{aligned}
\end{equation}
Consider the first term above $F_{\gamma_{N^{*},R}}^{}{(\gamma)}$, the exponential term can be decomposed using Taylor's series as
\begin{equation}
\begin{aligned}
F_{\gamma_{N^*,R}}^{}{(\gamma)}=&1-[1-\lambda_{\psi}\gamma+\frac{(\lambda_{\psi}\gamma)^{2}}{2!}-\frac{(\lambda_{\psi}\gamma)^{3}}{3!}\\&+\frac{(\lambda_{\psi}\gamma)^{4}}{4!}-\cdots],
\end{aligned}
\end{equation}
where $\lambda_{\psi}=\frac{1}{\overbar{\gamma}_{\psi}}$.

\newpage
\begin{strip}
\begin{equation}
\begin{aligned}
&ASEP_{tot}= \frac{a}{2}\Bigg[1-\sqrt\frac{b}{\Lambda_{1}}+K{K-1\choose N-1} \sum_{k=0}^{K-N} {K-N\choose k}{\frac{(-1)^{k}}{(k+N)}}\bigg\{\sqrt\frac{b}{\Lambda_{1}} -\sqrt\frac{b}{\Lambda_{2}}-\sqrt\frac{b}{\pi}\Bigg\langle\Lambda_{1}^{-\frac{1}{2}}\\&\times\large{G}_{2,2}^{1,2} \left[ \mu({\frac{\Lambda_{1}^{-\frac{1}{2}}}{\overbar{\gamma}_{s,r}}}) \left|
\begin{array}{cc} \frac{1}{2},1 \\ \mu,0 \end{array} \right. \right]
+\Lambda_{2}^{-\frac{1}{2}}\large{G}_{2,2}^{1,2} \left[ \mu({\frac{\Lambda_{2}^{-\frac{1}{2}}}{\overbar{\gamma}_{s,r}}}) \left|
\begin{array}{cc} \frac{1}{2},1 \\ \mu,0 \end{array} \right. \right]-\frac{\Lambda_{1}^{-\frac{1}{2}}}{\Gamma{(\mu)}}\large{G}_{2,2}^{1,2} \left[ \mu({\frac{\Lambda_{1}^{-\frac{1}{2}}}{\overbar{\gamma}_{r,s}}}) \left|
\begin{array}{cc} \frac{1}{2},1 \\ \mu,0 \end{array} \right. \right]+\frac{\Lambda_{2}^{-\frac{1}{2}}}{\Gamma{(\mu)}}\\&\times\large{G}_{2,2}^{1,2} \left[ \mu({\frac{\Lambda_{2}^{-\frac{1}{2}}}{\overbar{\gamma}_{r,s}}}) \left|
\begin{array}{cc} \frac{1}{2},1 \\ \mu,0 \end{array} \right. \right]\Bigg\rangle  \bigg\}
+\frac{\sqrt{b}}{\Gamma{(\mu)}\sqrt{\pi}}\Bigg\langle\Lambda_{1}^{-\frac{1}{2}}\large{G}_{2,2}^{1,2} \left[ \mu({\frac{\Lambda_{1}^{-\frac{1}{2}}}{\overbar{\gamma}_{s,r}}}) \left|
\begin{array}{cc} \frac{1}{2},1 \\ \mu,0 \end{array} \right. \right]+\Lambda_{1}^{-\frac{1}{2}}\large{G}_{2,2}^{1,2} \left[ \mu({\frac{\Lambda_{1}^{-\frac{1}{2}}}{\overbar{\gamma}_{r,s}}}) \left|
\begin{array}{cc} \frac{1}{2},1 \\ \mu,0 \end{array} \right. \right]\Bigg\rangle\\&-\frac{\sqrt{b}}{\Gamma{(\mu)}^{2}\sqrt{\pi}}\Lambda_{1}^{-\frac{1}{2}}
G_{0,1:2,2:2,2}^{1,0:1,2:1,2}\left[\begin{matrix}-\cr \frac{1}{2}\end{matrix}\Bigg\vert\begin{matrix}\frac{1}{2},1\cr \mu,0\end{matrix}\Bigg\vert\begin{matrix}\frac{1}{2}, 1\cr \mu,0\end{matrix}\Bigg\vert \frac{\mu}{\Lambda_{1}\overbar{\gamma}_{s,r}},\frac{\mu}{\Lambda_{1}\overbar{\gamma}_{r,s}}\right] + \frac{K\sqrt{b}}{\Gamma{(\mu)}^{2}\sqrt{\pi}}{K-1\choose N-1} \sum_{k=0}^{K-N} {K-N\choose k}{\frac{(-1)^{k}}{(k+N)}}\\&\times\bigg\{\Lambda_{1}^{-\frac{1}{2}}
G_{0,1:2,2:2,2}^{1,0:1,2:1,2}\left[\begin{matrix}-\cr \frac{1}{2}\end{matrix}\Bigg\vert\begin{matrix}\frac{1}{2},1\cr \mu,0\end{matrix}\Bigg\vert\begin{matrix}\frac{1}{2}, 1\cr \mu,0\end{matrix}\Bigg\vert \frac{\mu}{\Lambda_{1}\overbar{\gamma}_{s,r}},\frac{\mu}{\Lambda_{1}\overbar{\gamma}_{r,s}}\right] - \Lambda_{2}^{-\frac{1}{2}}
G_{0,1:2,2:2,2}^{1,0:1,2:1,2}\left[\begin{matrix}-\cr \frac{1}{2}\end{matrix}\Bigg\vert\begin{matrix}\frac{1}{2},1\cr \mu,0\end{matrix}\Bigg\vert\begin{matrix}\frac{1}{2}, 1\cr \mu,0\end{matrix}\Bigg\vert \frac{\mu}{\Lambda_{2}\overbar{\gamma}_{s,r}},\frac{\mu}{\Lambda_{2}\overbar{\gamma}_{r,s}}\right]\bigg\}  \Bigg],
\end{aligned}
\end{equation}
\noindent\makebox[\linewidth]{\rule{18cm}{0.4pt}}
\end{strip}

 Assuming that $\overbar{\gamma}_{\psi}\rightarrow\infty$, the CDF can be reduced to $\lambda_{\psi}\gamma$ while truncated all the other terms. Therefore, the asymptotic PDF is simplified to $f_{\gamma_{N^*,R}}^{\infty}(\gamma)\approx\lambda_{\psi}$. Using the CDF and PDF approximations to get the asymptotic CDF, we obtain 
\begin{equation}
\begin{aligned}
F_{\gamma_{N^*,R}}^{}{(\gamma)}\simeq K{K-1\choose N-1}\frac{(\lambda_{\psi}\gamma)^{K-N+1}}{K-N+1}.
\end{aligned}
\end{equation}
The approximation of the FSO link is derived using the generalized incomplete gamma function expansion series~\cite{22}. Hence, the term $F_{\gamma_{S,R}}^{FSO/RF}{(\gamma)}$ is given by
\begin{equation}
\begin{aligned}
F_{\gamma_{X,Y}}^{FSO/RF}{(\gamma)}\simeq\frac{\Gamma{(\mu,0)}}{\Gamma{(\mu)}}+\frac{(\lambda_{\psi}\gamma)^{\frac{\alpha\mu}{2}}}{\mu\Gamma{(\mu)}}-1,
\end{aligned}
\end{equation}
where $\Gamma{(a,b)}$ is the generalized incomplete gamma function and $\gamma_{X,Y}$ represents the SNR for either $S \to R$ (or $R \to S$) links~\cite{22}. The downlink RF approximation for the link $R\rightarrow N^*$ is based on Taylor's series expansion, therefore $F_{\gamma_{R,N^{*}}}^{}{(\gamma)}$ is approximated as
\begin{equation}
\begin{aligned}
F_{\gamma_{R,N^*}}^{}{(\gamma)}\simeq\gamma}{\lambda_{\psi}.
\end{aligned}
\end{equation}
Finally, upon substituting (36), (37), and (38) into (33) and replacing $\gamma$ with $\gamma_{th}$, the approximated outage expression is obtained at high SNR with straightforward mathematical manipulation as
\begin{equation}
\begin{aligned}
P_{out}^{Tot\to\infty}&=\Psi_{1}(\lambda_{\psi}\gamma_{th})^{K-N+1}+2\big\{\Psi_{2}+\frac{(\lambda_{\psi}\gamma_{th})^{\frac{\alpha\mu}{2}}}{\mu\Gamma{(\mu)}}\big\}
\\&+\gamma_{th}}{\lambda_{\psi},
\end{aligned}
\end{equation}
where $\Psi_{1}={K-1\choose N-1}\frac{K}{K-N+1}$ and $\Psi_{2}=\frac{\Gamma{(\mu,0)}}{\Gamma{(\mu)}}-1$. Rewriting (39) in the form of $P_{out}^{Tot}\simeq G_{c}(SNR)^{-G_{d}}$, we obtain
\begin{equation}
\begin{aligned}
P_{out}^{Tot\to\infty}&=(\Upsilon_{1}{\frac{\hat\gamma_{\psi}}{\gamma_{th}})^{-(K-N+1)}}+(\Upsilon_{2}{{\frac{\hat\gamma_{\psi}}{\gamma_{th}})^{-(\frac{\alpha\mu}{2})}}}+({\frac{\hat\gamma_{\psi}}{\gamma_{th}}})^{-1},
\end{aligned}
\end{equation}
where the term $\Psi_{2}$ can be neglected, while $\Upsilon_{1}=\Psi_{1}^{-\frac{1}{(K-N+1)}}$ and $\Upsilon_{2}=({\frac{1}{2}\mu\Gamma{(\mu)})}^{-\frac{2}{\alpha\mu}}$. Clearly, the overall performance of the system is going to be dominated by the worst CDF among the RF and FSO/RF links. As a result, the overall performance equals to the minimum of all CDFs among the links, i.e. $\min{(K-N+1, \frac{\alpha\mu}{2},1)}$. Coding gain and diversity order for various scenarios of the system model are shown in Table III where $T_{i},i=1,2,3$ denote the term number in (40).
\begin{table}[!t]
\caption{Coding gain and diversity order of the system model.}
\label{Table.3}
\centering
\small\addtolength{\tabcolsep}{0.7pt}
\begin{tabular}{ c c c } 
\hline
Domination links & Diversity order & Coding gain \\  [0.5ex]
\hline
$T_{1}$  & $K-N+1$ & $\frac{\Upsilon_{1}}{\gamma_{th}}$\\ 

$T_{2}$  & $\frac{\alpha\mu}{2}$ & $\frac{\Upsilon_{2}}{\gamma_{th}}$\\

$T_{3}$  & 1 & $\frac{1}{\gamma_{th}}$ \\

$T_{1}$ and $T_{3}$  & $K-N+1\simeq1$ & $\frac{\Upsilon_{1}}{\gamma_{th}}+\frac{1}{\gamma_{th}}$\\

$T_{1}$ and $T_{2}$  & $K-N+1\simeq\frac{\alpha\mu}{2}$ & $\frac{\Upsilon_{1}}{\gamma_{th}}+\frac{\Upsilon_{2}}{\gamma_{th}}$\\

$T_{2}$ and $T_{3}$  & $\frac{\alpha\mu}{2}\simeq1$ & $\frac{\Upsilon_{2}}{\gamma_{th}}+\frac{1}{\gamma_{th}}$\\

$T_{1}$, $T_{2}$ and $T_{3}$  & $K-N+1\simeq\frac{\alpha\mu}{2}\simeq1$ & $\frac{\Upsilon_{1}}{\gamma_{th}}+\frac{\Upsilon_{2}}{\gamma_{th}}+\frac{1}{\gamma_{th}}$\\
\hline
\end{tabular}
\end{table}

\section{Simulation and Numerical Results}
\vspace*{10pt}

In this section, outage probability and ASEPs of the system model are analyzed using Monte-Carlo simulations to verify the derived analytical expressions and asymptotic approximations. Furthermore, we investigate the effect of various parameters such as number of nodes $K$ on opportunistic scheduling scheme and asymptotic approximation, and the values of $\alpha$-$\mu$ fading model on the overall system performance. We assume in our simulations i.i.d. channel fading coefficients for all links, perfect pointing or negligible pointing error between the transmitter and receiver antennas, same $\alpha$-$\mu$ parameters for both $S \to R$ and $R \to S$ links, and the transmitted power by all terminals in the system over RF and/or FSO links are equally distributed among them such that $P_{S}^{}=P_{R}^{}=P_{N_{k}}^{}=\frac{1}{3}P_{T}$ where $P_{T}$ is the maximum power budget of the system.\par
\begin{figure}[!t]
\centering
\captionsetup{justification=centering}
\includegraphics[scale=0.48]{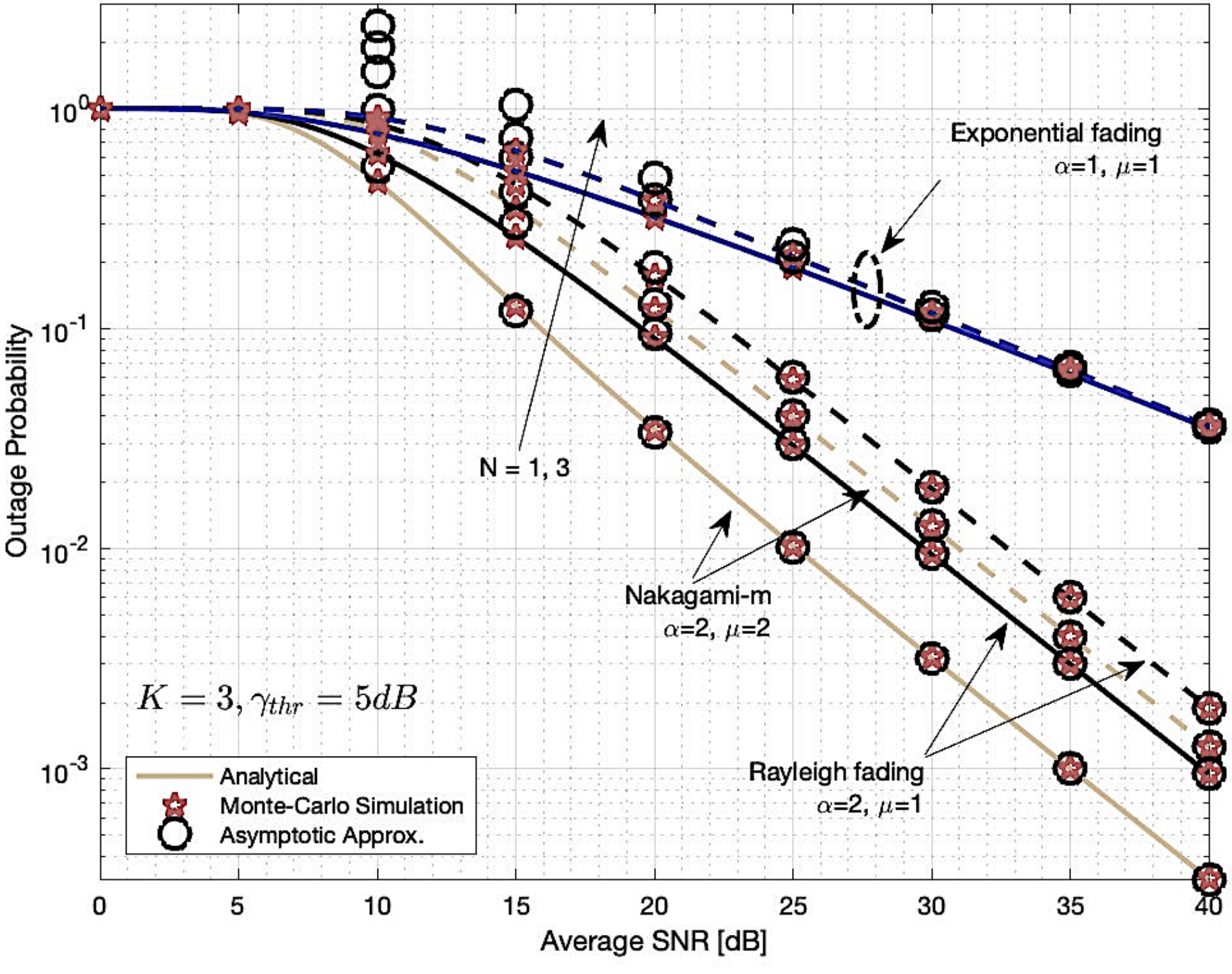}
\caption{Outage performance over fading channels for opportunistic scheduling using $N$ nodes.}
\label{Fig.3}
\end{figure}
Considering the RF/RF system model, Fig. 3 shows the impact of the opportunistic scheduling on the overall outage probability using $N$ out of $K$ nodes over Nakagami-$m$ ($\alpha=2$, $\mu=2$), Exponential ($\alpha=1$, $\mu=1$), and Rayleigh ($\alpha=2$, $\mu=1$) fading channels . We observe an exact match between our derived expressions and Monte-Carlo simulation, and close asymptotic approximation at high SNR. Obviously, the performance of Nakagami-$m$ is better than Exponential and Rayleigh since it is used to model line-of-sight (LOS) scenarios over the non line-of-sight (NLOS) for Rayleigh fading. However, we notice degradation in performance for the case of Nakagami-$m$ compared to both Exponential and Rayleigh channels of almost 5 dB coding loss as $N$ goes from 1 to 3. Also, we observe that the backup RF of $S\ \to R$, $R \to S$,  and $R \to N^*$ links are nearly dominating the overall outage performance over the link $N^* \to R$. This is due to the similar diversity orders of both links, i.e. $(G_{d}^{T_{2}}\simeq G_{d}^{T_{3}})<G_{d}^{T_{1}}$, where $G_{d}$ depends on $\alpha$ and $\mu$. However, the coding gain $G_{c}$ is affected when we apply opportunistic scheduling/selection among the $K$ nodes by the relay $R$. The diversity order of first RF link $G_{d}^{T_{1}}=G_{d}^{T_{3}}=1$ as $N\to3$ and diversity order increase but higher coding loss is encountered, i.e. $-G_{c}^{T_{1}+T_{3}}$. In the Exponential channel, the link between $S$ and $R$ is dominates the overall outage performance which results in the lowest coding loss $-G_{c}^{T_{2}}$.\par

\begin{figure}[!t]
\centering
\captionsetup{justification=centering}
\includegraphics[scale=0.48]{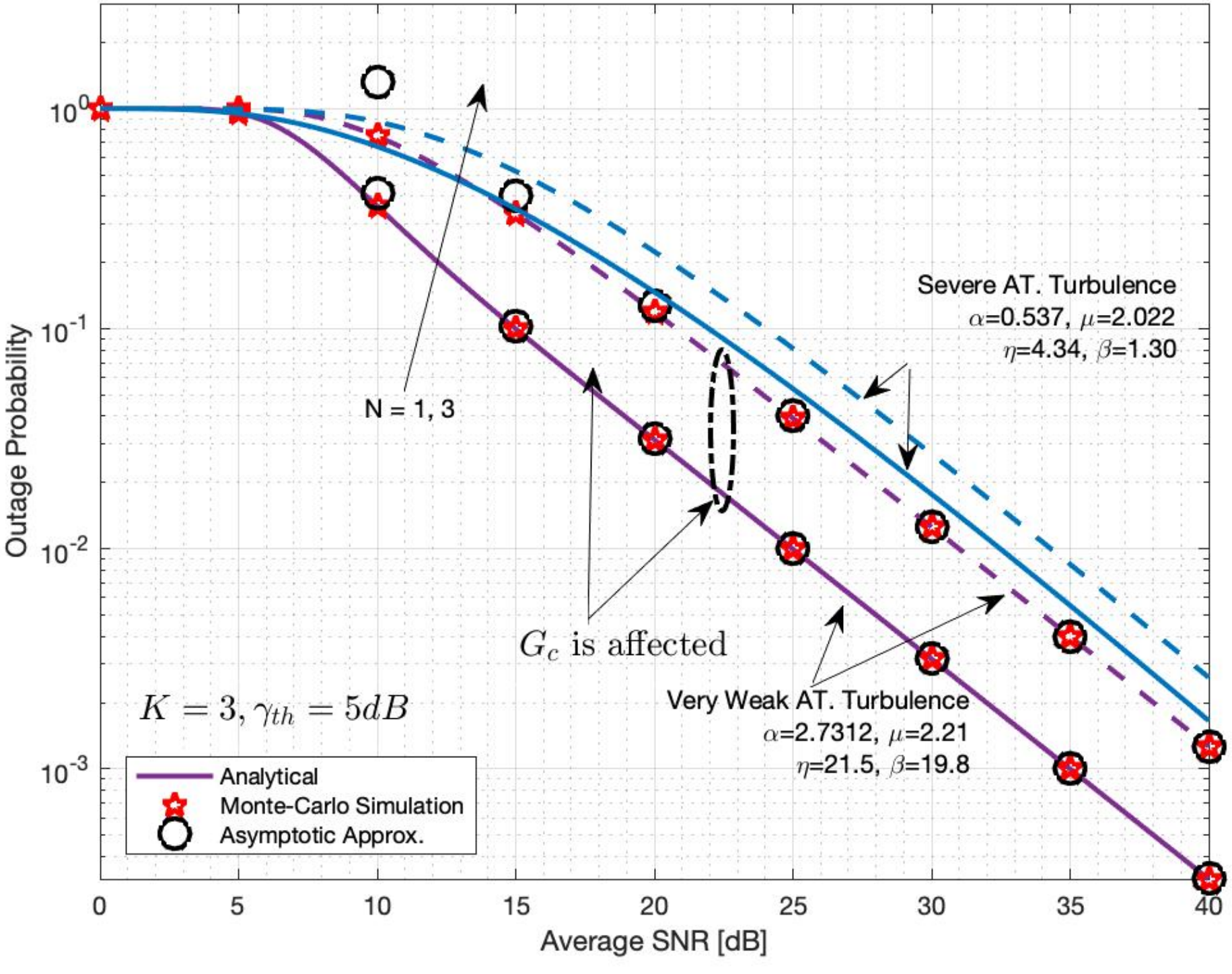}
\caption{Outage performance over weak and severe atmospheric turbulence under various opportunistic scheduling.}
\label{Fig.5}
\end{figure}

The outage probability of the RF/FSO system model is shown in Fig. 4 where an excellent match between the outage probability derived expressions with Monte-Carlo simulation for very weak atmospheric turbulence $(\alpha=2.7312,\mu=2.21)$. We notice an improved performance when the atmospheric turbulence is very weak compared to the severe situation. However, as the relay $R$ uses opportunistic scheduling among $K$ nodes, the outage performance degrades in the very weak atmospheric turbulence to around 6 dB coding loss in contrast to the severe case of almost 2 dB where FSO link dominates. An interesting point observed is that at very weak atmospheric turbulence, both RF links $R\to N^*$ and $N^*\to R$ dominate the outage performance of the system as $N\to3$ due to lower diversity achieved in these links than FSO link, i.e. $G_{d}^{T_{3}}/G_{d}^{T_{1}}< G_{d}^{T_{2}}$ and hence, $G_{d}^{T_{1}}=G_{d}^{T_{3}}=1$. On the other hand, the FSO link dominates the outage performance at severe atmospheric turbulence $(\alpha=0.579,\mu=2.022)$ because $G_{d}^{T_{2}}< G_{d}^{T_{1}}/G_{d}^{T_{3}}$ and lowest coding gain is achieved as given in Table III.\par



\begin{figure}[!b]
\centering
\captionsetup{justification=centering}
\includegraphics[scale=0.48]{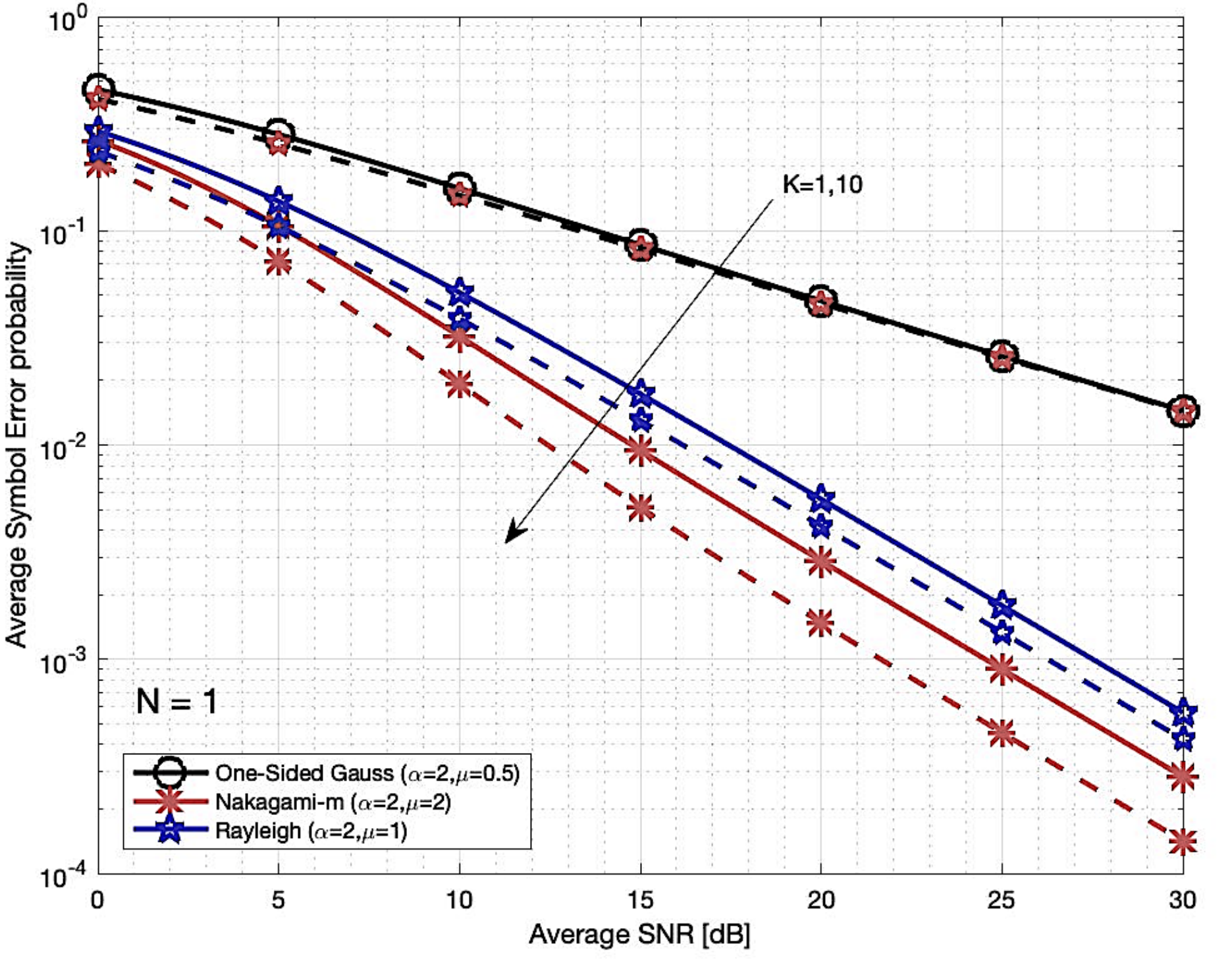}
\caption{ASEP over various RF fading channels and with different number of nodes $K$.}
\label{Fig.7}
\end{figure}

The average symbol error probability versus average SNR $(\overbar{\gamma}_{\psi})$ is simulated in Fig. 5  under best scheduling/selection by the relay $R$. We will investigate the impact of increasing the number of nodes $K$ over RF/RF Model. This gives insight on the overall ASEP performance of various RF fading channels where we notice the worst performance of One-Sided Gaussian over Nakagami-$m$ and Rayleigh fading channels. An improvement of around 2 dB coding gain occurs when $K$ increases from $1$ to $10$ for the Nakagami-$m$, and 1 dB coding gain for the Rayleigh channel while negligible improvement for One-Sided Gaussian fading channel. For instance, the $R\to N^{*}$ and link between $S$ and $R$ dominate the system performance over Nakagami-$m$ channel, and hence higher coding gain is achieved in ASEP. Conversely, low coding gain is attained for the One-Sided Gaussian channel due to the small coding gain value of link between $S$ and $R$.

The effect of opportunistic scheduling using $N$ nodes over various fading channels on the overall ASEP for fixed $K$ is shown in Fig. 6. We notice high coding loss in ASEP over Nakagami-$m$ compared to the Rayleigh and One-Sided Gaussian as $N$ increases. In One-Sided Gaussian, the backup RF link between $S$ and $R$ dominates the overall ASEP regardless of applied opportunistic scheduling at relay $R$. This is due to the lower diversity order in the link between $S$ and $R$ compared to all other links. However, the uplink and backup RF links are dominating the overall ASEP, and this degrades the performance for Nakagami-$m$ or Rayleigh.\par

\begin{figure}[!t]
\centering
\captionsetup{justification=centering}
\includegraphics[scale=0.48]{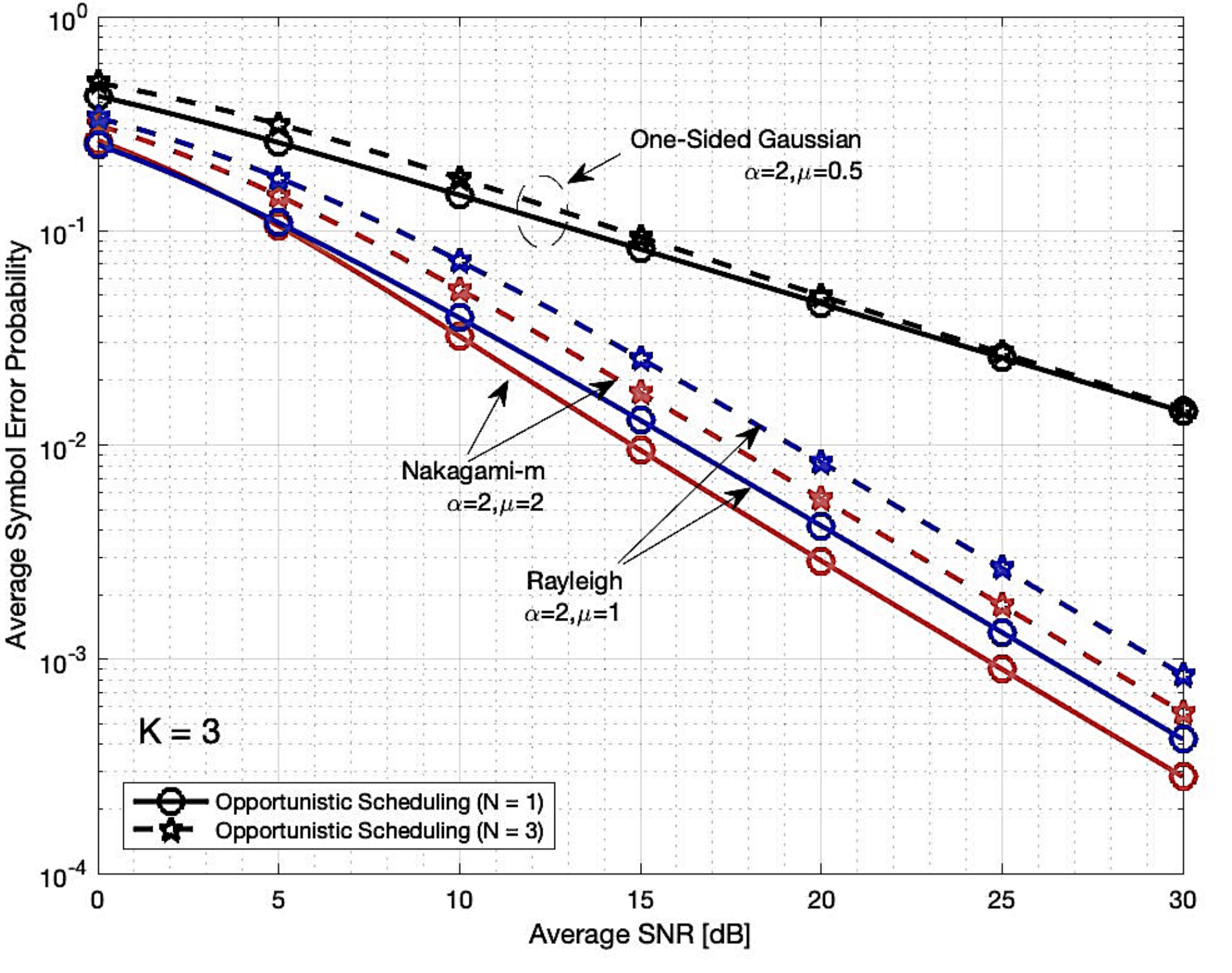}
\caption{ASEP over various RF fading channels and different number of nodes $N$ in opportunistic scheduling.}
\label{Fig.8}
\end{figure}

In the RF/FSO model, we investigate the ASEP over very weak and weak atmospheric turbulence conditions. In Fig. 7, the ASEP versus $\overbar{\gamma}_{\psi}$ under the assumption of best selection among the $K$ nodes is analyzed. We notice that in both very weak ($\alpha=2.73,\mu=2.21$) and weak ($\alpha=2.00,\mu=1.37$) atmospheric turbulence conditions, the RF and FSO links are dominating the overall ASEP due to the similar diversity order for both links and hence, an improvement of 2-3 dB coding gain is achieved as $K$ increases from 1 to 5.\par 

\begin{figure}[!t]
\centering
\captionsetup{justification=centering}
\includegraphics[scale=0.48]{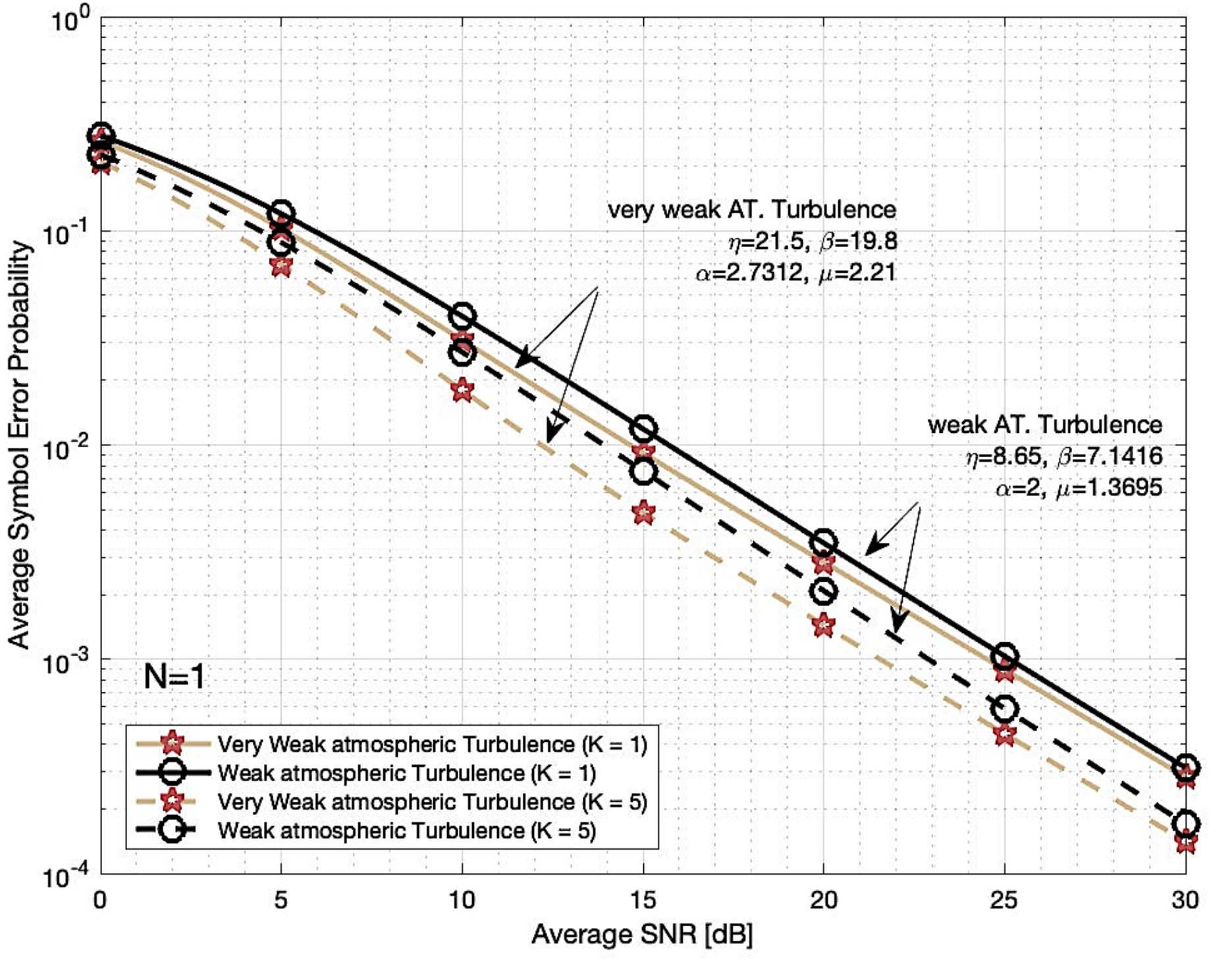}
\caption{ASEP for very weak and weak atmospheric turbulence cases with different number of nodes $K$.}
\label{Fig.9}
\end{figure}



The outage probability versus number of nodes $K$ is depicted in Fig. 8. It illustrates the achieved coding gain $G_{c}$ of both systems based on the different dominating conditions of all links as we increase number of nodes $K$ in the RF links between $N^*$ and $R$. We observe higher coding gain in very weak atmospheric turbulence $(\alpha=2.73,\mu=2.21)$ when $K$ increases from 1 to 5 compared to the severe case $(\alpha=0.537,\mu=2.022)$ in the FSO model while Nakagami-$m$ has the highest coding gain compared to all other fading channels in the backup RF/RF system. After $K=5$, no coding gain is attained in the outage performance for almost all fading distributions because as $K$ increases, the coding gain in the link between $N^*$ and $R$ goes to zero $G_{c}^{T_{1}}\to 0$ due to increase in the denominator of $T_{1}$ exponent. The coding gain $G_{c}$ can be deduced from Table III for all domination scenarios.

\begin{figure}[!b]
\centering
\captionsetup{justification=centering}
\includegraphics[scale=0.48]{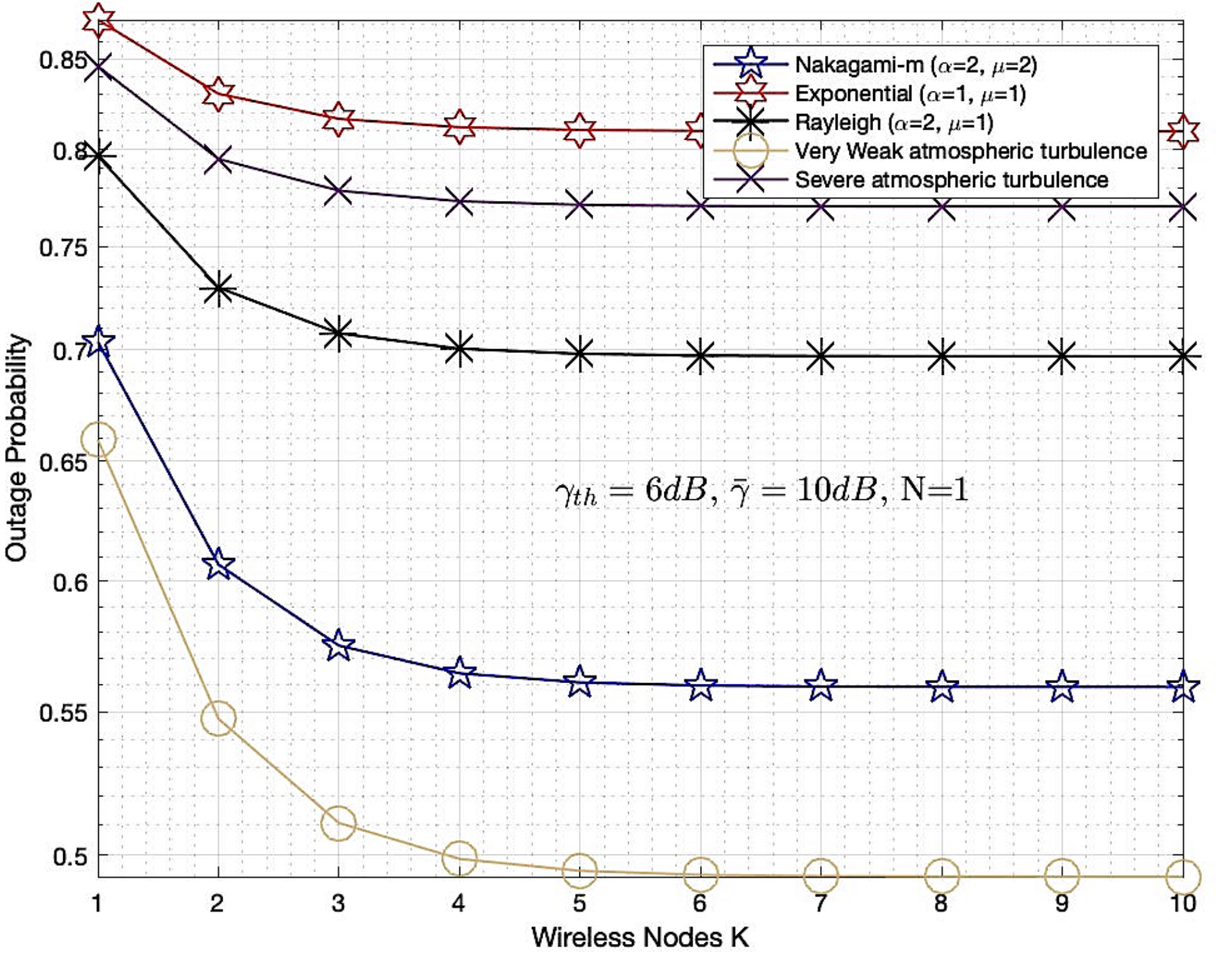}
\caption{Outage probability for $K$ wireless nodes over various fading channels.}
\label{Fig.11}
\end{figure}

\section{Conclusion}
The performance of hybrid RF/FSO with backup RF link, generalized opportunistic scheduling, and two-way DF relay over generalized $\alpha$-$\mu$ channel fading model was investigated. Outage probability and ASEP with asymptotic approximations were derived in closed-form and corroborated by Monte-Carlo simulations. Moreover, we observed the effect of number of nodes $K$ and opportunistic scheduling using $N$ nodes on the overall outage and ASEP for both mixed RF/FSO with backup RF/RF scenarios. The results show that hybrid/mixed two-way relaying has a good potential for next generation such as 5G specially for high data rate applications with reliable backhaul links. Due to the random behavior of atmospheric turbulence and pointing alignment error, our proposed hybrid RF/FSO can make the system more resilient in such unexpected weather conditions.

\section*{Acknowledgment}
The  authors  acknowledge  King  Fahd  University  of  Petroleum  and  Minerals  (KFUPM)  for supporting this research.

\ifCLASSOPTIONcaptionsoff
  \newpage
\fi


\begin{thebibliography}{99}

\bibitem{0}
M. A. Khalighi and M. Uysal,
``Survey on free space optical communication: a communication theory perspective," \textit{IEEE Commun. Surv. Tutorials}. vol. 16, no. 4, pp. 2231-2258, 2014.
\bibitem{1}
 L. C. Andrews, R. L. Phillips, and C. Y. Hopen, \textit{Laser Beam Scintillation with Applications}. Bellingham, WA: SPIE Press, 2001.
\bibitem{2}
E. Lee, J. Park, D. Han, and G. Yoon,
``Performance analysis of the asymmetric dual-hop relay transmission with mixed RF/FSO links," \textit{IEEE Photonics Technol. Lett.}, vol. 23, no. 21, pp. 1642-1644, Nov. 2011.
\bibitem{3}
I. S. Ansari, F. Yilmaz, and M. S. Alouini, 
``Impact of pointing errors on the performance of mixed RF/FSO dual-hop transmission systems", \textit{IEEE Wireless Commun. Lett.}, vol. 2, no. 3, pp. 351-354, Jun. 2013.
\bibitem{4}
Lei Kong, Wei Xu, Hua Zhang, and C. Zhao,
``Mixed RF/FSO two-way relaying system under generalized FSO channel with pointing error," \textit{Int'l
Conf. Ubiquitous and Future Net. (ICUFN’18)}, Vienna, Austria, pp. 264-269, May 2016.
\bibitem{5}
P. K. Sharma, A. Bansal, and P. Garg,
``Relay assisted bi-directional communication in generalized turbulence fading," \textit{IEEE/OSA J. Lightw. Technol.}, vol. 33, no. 1, pp. 133-139, Jan. 2015.
\bibitem{6}
N. Miridakis, M. Matthaiou, and G. Karagiannidis,
``Multiuser relaying over mixed RF/FSO links," \textit{IEEE Trans.Commun.}, vol. 62, no. 5, pp. 1634-1645, Mar. 2014
\bibitem{7}
A. M. Salhab,
``Performance of multiuser mixed RF/FSO relay networks with generalized order user scheduling and outdated channel information," \textit{Arabian J. Sci. Eng.}, vol. 40, no. 9, pp. 2671-2683, Jul. 2015.
\bibitem{8}
Y. F. Al-Eryani, A. M. Salhab, and M. S. Alouini,
``Two-way multiuser mixed RF/FSO relaying: performance analysis and power allocation," \textit{IEEE/OSA J. Opt. Commun.},  vol. 10, no. 4, pp. 396-408, April. 2018.
\bibitem{9}
J. H. Churnside and S. F. Clifford, ``Log-normal Rician probability density
function of optical scintillations in the turbulent atmosphere,” \textit{J.
Opt. Soc. Am. A}, vol. 4, pp. 1923-1930, Oct. 1987.
\bibitem{10}
 W. Huang, J. Takayanagi, T. Sakanaka, and M. Nakagawa,
``Atmospheric optical communication system using subcarrier PSK modulation," \textit{IEICE Trans. Commun.}, vol. E76-B, no. 9, pp. 1169-1177, 1993.
\bibitem{11}
M. K. Simon and M. S. Alouini,
\textit{Digital Communication over Fading Channels,} 2nd ed. Hoboken, New Jersey: Wiley, 2005.
\bibitem{12}
M. D. Yacoub, ``The $\alpha$-$\mu$ distribution: A physical fading model for the
Stacy distribution," \textit{IEEE Trans. Veh. Technol.}, vol. 56, no. 1, pp. 27–34,
Jan. 2007.
\bibitem{13}
A. M. Magableh and M. M. Matalgah,
``Moment generating function of the generalized $\alpha$-$\mu$ distribution with applications," \textit{IEEE Commun. Lett.}, vol. 13, no. 6, pp. 411-413, June 2009.
\bibitem{14}
I. S. Gradshteyn and I. M. Ryzhik,
\textit{Table of integrals, series, and products,} 7th ed. San Diego, California: Academic, 2014.
\bibitem{16}
R. J. Vaughan and W. N. Venables, \textit{Permanent expressions for order
statistics densities,} J. Roy. Statist. Soc. Ser. B, vol. 34, 1972.
\bibitem{17}
Y. A. Brychkov, O. Marichev, and A. Prudnikov,
``Integrals and Series, vol 3: more special functions", 1986.
\bibitem{18}
 N. D. Chatzidiamantis and G. K. Karagiannidis, ``On the distribution of
the sum of Gamma-Gamma variates and applications in RF and optical
wireless communications," \textit{IEEE Trans. Commun.}, vol. 59, no. 5,
pp. 1298-1308, May 2011.
\bibitem{19}
M. R. McKay, A. L. Grant, and I. B. Collings,
``Performance analysis of MIMO-MRC in double-correlated Rayleigh environments," \textit{IEEE Trans. Commun.}, vol. 55, no. 3, pp. 497-507, Mar. 2007.
\bibitem{20}
Wolfram, 
``The Wolfram functions site", [Online], Available: http://functions.wolfram.com.
\bibitem{21}
I. S. Ansari, S. Al-Ahmadi, F. Yilmaz, M. S. Alouini, and H. Yanikomeroglu,
``A new formula for the BER of binary modulations with dual-branch selection over generalized-$K$ composite fading channels," \textit{IEEE Trans. Commun.}, vol. 59, no. 10, pp. 2654-2658, 2011.
\bibitem{22}
G. Nemes and A. O. Daalhuis,
``Asymptotic expansions for the incomplete gamma function in the transition regions," accepted for publication in \textit{Math. Comp.}, 2018.
\end{thebibliography}
\end{document}